\documentclass[a4paper,11pt,twoside]{article}

\usepackage{geometry} 
\usepackage{amsmath}
\usepackage{amssymb}
\usepackage{color}
\usepackage{graphicx}

\newcommand{\omi}[1]{}

\newlength{\myem}
\newcounter{mysubequation}[equation]

\makeatletter
\renewcommand{\section}{\@startsection{section}{1}{0em}%
        {-3.5ex \@plus -1ex \@minus -.2ex}%
        {2.3ex \@plus.2ex}%
        {\normalfont\large\bfseries}}
\renewcommand{\subsection}{\@startsection{subsection}{2}{0em}%
        {-3.25ex\@plus -1ex \@minus -.2ex}%
        {1.5ex \@plus .2ex}%
        {\normalfont\bfseries}}
\renewcommand{\subsubsection}%
        {\@startsection{subsubsection}{3}{0em}%
        {-3.25ex\@plus -1ex \@minus -.2ex}%
        {1.5ex \@plus .2ex}%
        {\normalfont\itshape}}
\makeatother

\numberwithin{equation}{section}
\newcommand{\be}[1]{\begin{equation} #1 \end{equation}}

\newcommand{\SISSA}{SISSA/ISAS and INFN, I--34013 Trieste, Italy}

\newcommand{\preprintnumber}{%
SISSA--24/2007/EP}
 
\newcommand{\titletext}{R-symmetry breaking, runaway directions and global 
symmetries in O'Raifeartaigh models} 
\newcommand{\authortext}{\large Luca Ferretti
\medskip\\\em\normalsize 
\SISSA
        }
\newcommand{\abstracttext}{We discuss O'Raifeartaigh models with general R-charge assignments, introduced by Shih to break R-symmetry spontaneously. We argue that most of these models have runaway directions related to the R-symmetry. In addition, we study  the simplest model with a U(N) global symmetry and show that in a range of parameters R-symmetry is spontaneously broken in a metastable vacuum.}

\title{
\normalsize
\hspace*{\fill}
\begin{tabular}[t]{l}\preprintnumber\end{tabular}
\vspace{3\baselineskip}\\\Large\bfseries\titletext\bigskip}
\author{\begin{minipage}[t]{0.8\textwidth}
\normalsize\centering\authortext
\end{minipage}}
\date{}

\begin{document}

\bigskip
\maketitle
\begin{abstract}\normalsize\noindent
\abstracttext
\end{abstract}\normalsize\vspace{\baselineskip}

\section{Introduction}
Recently O'Raifeartaigh models \cite{oraif} raised some interest as appealing candidates for the hidden sector of low-scale SUSY models. The main reason is the discovery of metastable SUSY-breaking vacua in $\mathcal{N}=1$ SQCD \cite{iss1}, that can be seen in the low-energy effective theory as vacua of an O'Raifeartaigh-type model. These models are perturbative and calculable and they therefore have the advantage that their properties can be reliably studied even if SUSY is spontaneously broken.

Most of the O'Raifeartaigh models featured in the literature have flat directions of SUSY-breaking vacua at the classical level, parametrized by fields $X_n$ of R-charge 2. Quantum corrections lift these vacua in such a way that
the true vacuum lies at $X_n=0$ and R-symmetry is unbroken. Shih noted that this is a consequence of the particular R-charge assignment of these models: R-symmetry is unbroken in models which only have fields with $R=0$ or $R=2$, whereas in models with more general R-charges there can be spontaneous R-symmetry breaking \cite{shih}. 

The simplest model which breaks R-symmetry spontaneously for some values of its parameters is:
\be{W=fX+\lambda X\phi_{(1)}\phi_{(-1)}+m_1 \phi_{(3)} \phi_{(-1)} +\frac{1}{2}m_2 \phi_{(1)}^2\label{worig}}
where $R(X)=2$ and $R(\phi_{(k)})=k$. Classically this model has a flat direction of local extrema given by $\phi_{(3)}=\phi_{(1)}=\phi_{(-1)}=0$; this direction is parametrized by $X$ with potential $V(X)=|f|^2$ and is a local minimum for $|X|<\frac{m_1^2m_2}{2\lambda^2 f}-\frac{f}{2m_2}$. 
Quantum corrections modify the tree-level potential as $V(X)=|f|^2+m_X^2 |X|^2+\ldots$ with $m_X^2<0$ in a large region of the space of couplings. In this case the potential $V(X)$ can have a (local) minimum away from the origin and the R-symmetry is broken in this vacuum.

An interesting observation is that the above vacuum is metastable because of the existence of a runaway direction \cite{shih}:
\be{\phi_{(1)}=-\frac{f}{\lambda \phi_{(-1)}},\ X=\frac{m_2 f}{\lambda^2 \phi_{(-1)}^2} ,\ \phi_{(3)}=\frac{m_2 f^2}{m_1\lambda^2\phi_{(-1)}^3},\ \phi_{(-1)}\rightarrow 0}
 The runaway behavior of this model has two  
 properties. First of all, the potential goes to zero along the runaway direction, therefore the runaway vacuum is supersymmetric. Secondly, the runaway direction can be seen as a rescaling of fields \be{\varphi(\epsilon) =\epsilon^{-R(\varphi)} \varphi(0) \quad,\quad \epsilon \rightarrow 0\label{resc}} 
 A natural question arises: is this behavior a feature of a large class of models with general R-charges, or does it happen only in this example?
 
 Another interesting observation is that this model cannot be extended with global symmetries under which the fields transform as complex representations\footnote{The mass term for $\phi_{(1)}$ requires that the representations $\mathcal{R}(\phi_{(1)})\otimes_{simm}\mathcal{R}(\phi_{(1)})\supset \mathbf{1}$, therefore $\mathcal{R}(\phi_{(1)})$ cannot be an irreducible complex representation; 
  the same is true for the other fields, because $\mathcal{R}(\phi_{(-1)})\otimes\mathcal{R}(\phi_{(1)})\supset \mathbf{1}$ and $\mathcal{R}(\phi_{(3)})\otimes\mathcal{R}(\phi_{(1)})\supset \mathbf{1}$.}. Global symmetries are interesting because they can play an important role in mediating supersymmetry breaking: for example, they can be gauged and communicate SUSY breaking directly through gauge interactions, as in \cite{med1,med2,ak} or through a messenger sector, as in \cite{mn1,as,mn2}. Non-abelian global symmetries can also be useful when looking for an ultraviolet completion of these models as effective theories of strongly-coupled gauge theories, as in \cite{iss1}. It is easy to write a model with real representations, for example $SO(N)$ fundamentals:
 \be{W=fX+\lambda X\phi_{(1)}^\alpha\phi_{(-1)}^\alpha+m_1 \phi_{(3)}^\alpha \phi_{(-1)}^\alpha +\frac{1}{2}m_2 \phi_{(1)}^\alpha \phi_{(1)}^\alpha}
 or to add other fields which interact only with $X$ and play no role in breaking SUSY:
\be{\Delta W=\lambda' X \bar\varphi_\alpha \varphi^\alpha}
but it could be interesting to have an O'Raifeartaigh model with spontaneous R-symmetry breaking where the SUSY-breaking sector contains fields in complex representations of a flavour symmetry.
 
In this paper we address both these issues. We argue that many O'Raifeartaigh models with general R-charge assignment have runaway behavior. Runaway directions in these models are related to the R-symmetry of the theory, as in the above example. We also study the simplest model with spontaneous R-symmetry breaking which contains fields in the fundamental and antifundamental representations of U(N). Several appendixes contain those explicit computations and/or numerical
simulations which were omitted from the main text not to break the logical
flow.

In section \ref{simple} we prove the existence of runaway directions in a simple class of  models 
with a single pseudomodulus $X$, analyzed in \cite{shih}. We show that in all models with a field with $R \neq 0,1,2$ there is a runaway direction and that the potential goes to zero along this direction, therefore the runaway vacuum is supersymmetric. In section \ref{more} we discuss the case of more pseudomoduli. We show examples of models with SUSY and non-SUSY runaway vacua and we argue that most of the models in this class have runaway directions. In section \ref{general} we end with some  
considerations about runaway vacua in general O'Raifeartaigh models. 
In section \ref{flavour} we study the simplest model which contains fields in complex representations of a flavour symmetry. We show that R-symmetry is spontaneously broken in this model for some range of parameters. The R-breaking vacua are always metastable.

In appendix \ref{more2} we extend the analysis of spontaneous R-symmetry breaking 
to the case of models with more pseudomoduli. In appendix \ref{proof} we complete the proof of section \ref{simple} on the existence of runaway directions. In appendix \ref{cond} we discuss some sufficient conditions for runaway directions in models with more pseudomoduli. Finally, in appendix \ref{vacua} we explain the relation between runaway directions and the  issues of metastability discussed recently in the literature.

\section{Runaway directions}
\subsection{Models with a single pseudomodulus}\label{simple}
We consider a simple class of models considered in \cite{shih}. These models are generalizations of the model (\ref{worig}): they consist of a chiral superfield $X$ with $R(X)=2$ and $n_\phi$ chiral superfields $\phi_i$. All these fields have a canonical K\"ahler potential and a superpotential
\be{W=fX+\frac{1}{2}(M^{ij}+N^{ij}X)\phi_i \phi_j \label{wshih}}
where $M,N$ are symmetric complex matrices with $\det (M)\neq 0$. Note that the last condition constrains both the R-charges and the field content of the model; for example, it implies that the number of fields with $R=r$ is the same as the number of fields with $R=2-r$. Moreover, R-symmetry constrains the possible nonzero entries in these matrices: \be{M^{ij}\neq 0 \Rightarrow R(\phi_i)+R(\phi_j)=2 \quad ,\quad N^{ij}\neq 0 \Rightarrow R(\phi_i)+R(\phi_j)=0\label{constr}}
 Apart from these restrictions and those coming from other symmetries, we consider $M,N$ to be generic. 

According to general arguments, R-symmetry implies that this superpotential can break SUSY\cite{ns}. In fact, it is shown in \cite{shih} that SUSY is always broken in these models. It is also shown that a necessary condition for having R-symmetry breaking
vacua is that fields with R-charges different from 0 and 2 exist, and it is
argued that this is also sufficient for a wide range of parameters. 

Let's review the argument
for SUSY breaking in these models. The equations for a SUSY vacuum $\partial_a W=0$ are
\begin{eqnarray}
f+\frac{1}{2}N^{ij}\phi_i\phi_j=0 \label{ex}\\
(M^{ij}+N^{ij}X)\phi_j=0\label{ea}
\end{eqnarray}
and cannot be solved simultaneously. To prove this it is sufficient to note that if $\det(M+NX)\neq0$ the only solution for (\ref{ea}) is $\phi_i=0$ which cannot satisfy (\ref{ex}). It can be shown that $\det(M+NX)=\det(M)$ if R-symmetry is required, and SUSY is therefore broken in all models with $\det (M)\neq 0$. However, this argument only refers to finite values of the fields and does not exclude a supersymmetric runaway vacuum.

To obtain a SUSY runaway vacuum, we classify the equations (\ref{ea}) according to their R-charge:
\begin{eqnarray}
(M^{ij}+N^{ij}X)\phi_j=0\quad &,&\quad R(\phi_i)< 2\label{ea+}\\
(M^{kj}+N^{kj}X)\phi_j=0\quad &,&\quad R(\phi_k)= 2\label{ea0}\\
(M^{mj}+N^{mj}X)\phi_j=0\quad &,&\quad R(\phi_m)> 2\label{ea-}
\end{eqnarray}
The equations (\ref{ea+}),(\ref{ea0}),(\ref{ea-}) have R-charges positive, zero and negative respectively. Given the above argument, there is no solution for the system of equations (\ref{ex}),(\ref{ea+}),(\ref{ea0}),(\ref{ea-}). In fact the equations (\ref{ex}),(\ref{ea0}),(\ref{ea-}) are not compatible, because (\ref{ex}) requires at least one field with non-positive R-charge to be nonzero, while equations (\ref{ea0}),(\ref{ea-}) force all fields with non-positive R-charge to zero.
 However there could be a field configuration $X',\phi'_i$ which solves the subsystem (\ref{ex}),(\ref{ea+}),(\ref{ea0}). If this is the case, the potential of these fields is \be{V=\sum_{R(\phi_m)> 2}|(M^{mj}+N^{mj}X')\phi'_j|^2} and it goes to zero along the direction parametrized by $\epsilon$ in (\ref{resc}): 
\be{\phi_i(\epsilon) =\epsilon^{-R(\phi_i)} \phi'_i \quad,\quad 
X(\epsilon)=\epsilon^{-2} X' \quad, \quad
\epsilon \rightarrow 0\label{resc2}} 
This means that the theory cannot have a lower ground state, and there is a runaway direction parametrized by non-unitary R-symmetry transformations (\ref{resc2}).

In appendix \ref{proof} we prove that in this class of models it is always possible to solve (\ref{ex}),(\ref{ea+}),(\ref{ea0}) at the same time if there are fields with\footnote{This is not completely correct, because R-charge is defined only up to addition of other $U(1)$ charges. So a more correct formulation is: we can always solve (\ref{ex}),(\ref{ea+}),(\ref{ea0}) at the same time if  for every choice of R-charges there is at least a field with $R\neq~0,1,2$.} $R\neq~0,1,2$. 
For the models (\ref{wshih}) which satisfy this condition, this result implies that local minima of the potential always correspond to metastable vacua, and that the potential  shows a runaway behavior.
The properties of these models are therefore very different from usual O'Raifeartaigh models. 

Many models in this class have metastable R-breaking vacua. In fact the presence of fields with $R\neq~0,1,2$ in these models corresponds both to the necessary condition for spontaneous R-symmetry breaking and to the sufficient condition for runaway behavior.  
An interesting consequence is that for this class of models, spontaneous R-symmetry breaking implies metastability.

\subsection{Models with more pseudomoduli}\label{more}
To understand what can happen in more general models, we add to the previous models a set of fields $Y_a$ with $R(Y_a)=2$, canonical K\"ahler potential and superpotential
\be{W=fX+\frac{1}{2}(M^{ij}+N^{ij}X+Q^{ij}_aY_a)\phi_i \phi_j \label{wmore}}
where $Q_a$ are generic symmetric complex matrices with  \be{Q^{ij}_a\neq 0 \Rightarrow R(\phi_i)+R(\phi_j)=0\label{constr2}}
 Similarly to the previous case, these models breaks SUSY. 
The proof is identical to the previous one if we substitute $NX$ with $NX+Q_aY_a$, because it depends only on the properties (\ref{constr}),(\ref{constr2}). These models can also have R-symmetry breaking vacua for some values of parameters. This is obvious in the limit $Q_a\rightarrow 0$, where they reduce to the models (\ref{wshih}). An analysis of R-symmetry breaking in models with more pseudomoduli is presented in appendix \ref{more2}.  

The analysis of runaway directions is  different from the case with a single pseudomodulus.
To see the difference, we analyze some simple examples\footnote{Note that throughout this paper the indices in parentheses correspond to the R-charges of the fields.}:
\begin{itemize}
\item 
This is a simple modification of the Shih model (\ref{worig}) with a $Y$ field:
\be{W=fX+(\lambda X+\eta Y) \phi_{(1)}\phi_{(-1)}+m_1 \phi_{(3)} \phi_{(-1)} +\frac{1}{2}m_2 \phi_{(1)}^2\label{wy}}
Classically this model has flat directions of SUSY-breaking vacua with $\phi_{(3)}=\phi_{(1)}=\phi_{(-1)}=0$ for some range of parameters. These flat directions are parametrized by $X,Y$ and are lifted by quantum effects. As in the original model, the quantum vacuum can break the R-symmetry, depending on the choice of parameters. 

Here the equations $\partial_X W=0$, $\partial_Y W=0$ have $R=0$ but cannot be solved at the same time. This means that there are no SUSY runaway vacua.  However there is a runaway direction
\be{\phi_{(1)}=-\frac{f}{\lambda' \phi_{(-1)}} ,\ X+\frac{\eta}{\lambda}Y=\frac{m_2 f}{\lambda'^2 \phi_{(-1)}^2}  ,\ \phi_{(3)}=\frac{m_2 f^2}{m_1\lambda'^2\phi_{(-1)}^3},\ \phi_{(-1)}\rightarrow 0}
with $\lambda'=(|\lambda|^2+|\eta|^2)/\bar{\lambda}$. This non-SUSY runaway vacuum minimizes the potential and the above vacua are therefore metastable. 
\item This simple model has a $U(1)$ symmetry $\phi^\pm_{(k)}\rightarrow e^{\pm i\theta} \phi^\pm_{(k)}$ and shows a different behavior:
 \begin{eqnarray}W=fX+(\lambda_+ X +\eta_+ Y) \phi^+_{(1)}\phi^-_{(-1)}+(\lambda_- X +\eta_- Y) \phi^+_{(-1)}\phi^-_{(1)}+\nonumber\\ +m_3\phi^+_{(3)} \phi^-_{(-1)} +m_1 \phi^+_{(1)} \phi^-_{(1)}+m_{-1}\phi^+_{(-1)}\phi^-_{(3)} \label{wy2}
 \end{eqnarray}
Here we can solve all the equations with $R>0$
in terms of $\phi^+_{(-1)}$,$\phi^-_{(-1)}$,$X$,$Y$  as in the models of section \ref{simple}, obtaining $\phi^\pm_{(1)}=-(\lambda_\mp X +\eta_\mp Y) \phi^\pm_{(-1)}/m_1$. The equations with $R=0$ become 
 \begin{eqnarray} fm_1- \left[2\lambda_+ \lambda_- X +(\lambda_+\eta_- +\lambda_- \eta_+)Y\right] \phi^+_{(-1)}\phi^-_{(-1)}=0 \\
 \left[2\eta_+ \eta_- Y +(\lambda_+\eta_- +\lambda_- \eta_+)X\right] \phi^+_{(-1)}\phi^-_{(-1)}=0
 \end{eqnarray} and can be easily solved with $\phi^+_{(-1)}\phi^-_{(-1)} \neq 0$.
  Then there is a SUSY runaway vacuum which corresponds to a field rescaling  $\phi^+_{(-1)},\phi^-_{(-1)}\rightarrow0$.
\end{itemize}  

Let's analyze the general case. The equations for a SUSY vacuum are:
\begin{eqnarray}
f+\frac{1}{2}N^{ij}\phi_i\phi_j=0 \quad & &\label{mex}\\
\frac{1}{2}Q_a^{ij}\phi_i\phi_j=0\quad & & \label{mey}\\
(M^{ij}+N^{ij}X+Q^{ij}_aY_a)\phi_j=0\quad &,&\quad R(\phi_i)< 2\label{mea+}\\
(M^{kj}+N^{kj}X+Q^{kj}_aY_a)\phi_j=0\quad &,&\quad R(\phi_k)= 2\label{mea0}\\
(M^{mj}+N^{mj}X+Q^{mj}_aY_a)\phi_j=0\quad &,&\quad R(\phi_m)> 2\label{mea-}
\end{eqnarray}
As in the case with a single pseudomodulus, the equations (\ref{mex}),(\ref{mea0}),(\ref{mea-}) are not compatible. Then
there are three 
cases:
\begin{itemize}
\item[(a)] If we can solve all the equations with non-negative R-charge (\ref{mex}),(\ref{mey}),(\ref{mea+}),(\ref{mea0}) at the same time, we can then rescale the solution as in (\ref{resc}) and obtain a runaway direction. The runaway vacuum is supersymmetric, and therefore all other vacua, if any, are metastable. 

This is what happens in model (\ref{wy2}). 
This case often happens for small $n_Y$.

\item[(b)] If it is not possible to solve the equations (\ref{mex}),(\ref{mey}),(\ref{mea+}),(\ref{mea0}) for any choice of R-charges, we look for absolute minima $\varphi_a^{min}$ of the potential $V_{min}(\varphi)=\min(V_+(\varphi),V_-(\varphi))$ with respect to all fields and all choices of R-symmetries, where $V_+$ and $V_-$ are
\be{ V_+=\sum_{R(\varphi_a)\leq 2}|\partial_{\varphi_a} W|^2 \quad, \quad V_-=\sum_{R(\varphi_a)\geq 2}|\partial_{\varphi_a} W|^2}
 Now there are two possibilities:
 \begin{itemize}
 \item[(b1)]
If there are $\varphi_a^{min}$ which solve both (\ref{mea+}) and (\ref{mea-}), these are the true vacua of the model, 
with a flat direction parametrized by R-charge rescalings. 

This is what happens in original O'Raifeartaigh model and in all models with R=0,2. 

\item[(b2)] Suppose that the absolute minimum is at $V_+(\varphi_a^{min})$.
 If there are no field configurations $\varphi_a^{min}$ which solve (\ref{mea+}),(\ref{mea-}) but there is a $\varphi_a^{min}$ which only solves (\ref{mea+}), we can then rescale this solution as in (\ref{resc}) and obtain a runaway direction. The runaway vacuum is not supersymmetric but it corresponds to the true vacuum of the system and therefore all other vacua, if any, are metastable. The same if we exchange (\ref{mea+}) with (\ref{mea-}) and $V_+$ with $V_-$.

This is what happens in model (\ref{wy}). This case often happens for large $n_Y$.
\end{itemize}
 
\item[(c)] The last possibility is that absolute minima of $V_{min}$ do not solve  (\ref{mea+}) nor  (\ref{mea-}). In this case there are no general results, but there can be non-SUSY stable vacua or runaway directions, depending on the details of the models.  
\end{itemize}
A model can belong to one or another of the above cases, depending on its parameters and field content.

It is possible to find sufficient conditions for the existence of runaway directions which consider only the field content of the model. Roughly speaking, there are runaway directions if $n_Y \gtrsim n_\phi/2$ and there are SUSY runaway vacua if $n_Y\lesssim n_\phi/2$. There is a (small) window of models without runaway vacua, but these conditions imply that most of these models have runaway directions.
The precise conditions and their proofs can be found in appendix \ref{cond}. 

\subsection{General considerations}\label{general}

The interesting result of the previous sections is that many O'Raifeartaigh models have a runaway behavior. In this section we argue that this behavior is quite common in O'Raifeartaigh models with general R-charge assignments.

We briefly review the usual O'Raifeartaigh models in our approach. (For more details about these models, see the lectures \cite{is-lect}.)
The superpotential is
\be{W=\sum_n X_n g_n(\phi_i) \quad  (n=1\ldots n_X, i=1\ldots n_0)}
where $R(X_n)=2$, $R(\phi_i)=0$. These models break SUSY because the conditions $g_n(\phi_i)=0$ are generally not compatible if $n_X>n_0$. The fields $\phi_i$ are determined by minimization of $V=\sum_n |g_n(\phi_i)|^2$; this means that the equations $\sum_n X_n \partial_j g_n=0$ have at least a nonzero solution $X_n=g^*_n(\phi^*_i)$. Rescaling this solution with respect to the R-charges, we obtain a flat direction of minima\footnote{Actually there is a $(n_X-n_0)$-dimensional space of solutions. R-symmetry rescaling acts as a dilatation in this space.}.

When there are fields with $R\neq 0,1,2$ the picture changes completely. In this case  equations for SUSY vacua have generically $R>0,R=0,R<0$ and because of R-symmetry it is not possible to solve all these equations. However, it is sufficient to solve the equations with $R\geq 0$ and then rescale all fields as in (\ref{resc}) to obtain a runaway direction with $V\rightarrow 0$. 

If we consider a generic (possibly non-renormalizable) superpotential, the number of equations with $R\geq 0$ is usually smaller than the number of fields on which these equations depend, so they can be often solved. This means that runaway directions are common in these models, and that SUSY-breaking vacua of these models are generally metastable. We have seen in section \ref{simple} an interesting class of models which show this behavior.

It can also happen that only equations with $R \leq 0$ can be solved. This is not common in the models studied in the previous sections, but can happen in general models. An example which appeared early in the literature is the runaway model of \cite{witten}:
\be{W=fX+\alpha X^2\phi}
with $R(X)=2$, $R(\phi)=-2$. In this model there are no equations with $R<0$, so if we solve the $R=0$ equation $f+2\alpha X\phi=0$ and then rescale the fields as $\phi\rightarrow \epsilon^{-2} \phi,X\rightarrow \epsilon^2 X$ we find a runaway direction with $V(X,\phi)\rightarrow 0$ as $\epsilon\rightarrow 0$. This runaway vacuum is the only vacuum of this model.

As we have seen in section \ref{more}, general models can also have different behavior. For example there can be a runaway direction with $V\rightarrow V_\infty >0$. This happens when the equations with $R>0$ (or $R<0$) can be solved, but it is not possible to solve those with $R=0$ at the same time. This direction can be a runaway vacuum or not, depending on the model and its parameters. 
Other models can have stable SUSY-breaking vacua or flat directions, as the usual O'Raifeartaigh models.

It is interesting that a relation often exists between R-symmetry breaking and  metastability. In \cite{iss2} it is argued that metastability is a general feature of realistic models of SUSY breaking. In fact R-symmetry must be a good symmetry for the theory to break SUSY, but a small explicit R-symmetry breaking interaction is needed to give mass to the R-axion; this explicit breaking generically restores supersymmetry in vacua far away from the origin of field space. Near the origin, R-symmetry is an approximate symmetry and SUSY is spontaneously broken in a metastable vacuum. It is not clear if metastability in the models of  \cite{iss2} and in our models are related. Some hints in this direction are discussed in appendix \ref{vacua}, where it is shown that runaway directions are often remnants of supersymmetric vacua generated by (small) explicit R-breaking terms in the superpotential.

Besides, the fact that runaway directions appear naturally in many O'Raifeartaigh models 
leads to speculations about possible applications in cosmology.
In fact runaway fields could play the role of the inflaton field if they could be stabilized at large vevs.

\section{Models with global symmetries}\label{flavour}
The models discussed in the previous sections can have non-abelian global symmetries. However models which necessarily have a field with $R=0,1$ can have only fields in real representations. An example of such a case is this small modification of the original Shih model (\ref{worig}) where $\phi_{(-1)},\phi_{(1)},\phi_{(3)}$ are $SO(N)$ fundamentals: 
 \be{W=fX+\lambda X\phi_{(1)}^\alpha\phi_{(-1)}^\alpha+m_1 \phi_{(3)}^\alpha \phi_{(-1)}^\alpha +\frac{1}{2}m_2 \phi_{(1)}^\alpha \phi_{(1)}^\alpha}
By looking at the Coleman-Weinberg formula
\be{V_{eff}^{(1-loop)}=\frac{1}{64\pi^2}\mathrm{Tr}\left(\mathcal{M}_B^4\ln\frac{\mathcal{M}_B^2}{\Lambda^2}-\mathcal{M}_F^4\ln\frac{\mathcal{M}_F^2}{\Lambda^2}\right)\label{cw}}
it is easy to see that the effective potential is related to that of the original Shih model by $V_{eff}^{(1-loop)}(X)_{SO(N)}=NV_{eff}^{(1-loop)}(X)$. Then the analysis in \cite{shih} goes unchanged (except for the height of the potential barrier for the metastable vacuum, which is not relevant) and the model shows spontaneous non-hierarchical R-symmetry breaking in a metastable vacuum for a wide range of parameters. The flavour symmetry is unbroken in the metastable vacuum.

If we wish to introduce complex representations, we must consider models without $R=0,1$ fields. The simplest example is
\begin{eqnarray}
W&=&fX+XN_5\phi_{(5)}^\alpha\phi_{(-5)\alpha}+XN_3\phi_{(3)}^\alpha\phi_{(-3)\alpha}+ \nonumber \\ &&+M_7\phi_{(7)}^\alpha\phi_{(-5)\alpha}+M_5\phi_{(5)}^\alpha\phi_{(-3)\alpha}+M_3\phi_{(3)}^\alpha\phi_{(-1)\alpha\label{wf}}
\end{eqnarray}
where $\phi_{(7)},\phi_{(5)},\phi_{(3)}$ are fields in the fundamental representation of a $U(N)$ flavour symmetry and  $\phi_{(-5)},\phi_{(-3)},\phi_{(-1)}$ are in the antifundamental\footnote{For a generic representation $\mathcal{R}$ of a group $G$, the only modification is $V_{eff}^{(1-loop)}(X)_{\mathcal{R}(G)}=\mathrm{dim}(\mathcal{R}(G))V_{eff}^{(1-loop)}(X)$.}. Also in this case we have $V_{eff}^{(1-loop)}(X)_{U(N)}=NV_{eff}^{(1-loop)}(X)$, therefore all relevant properties can be found from the model without the flavour symmetry:
\begin{eqnarray}
W&=&fX+XN_5\phi_{(5)}\phi_{(-5)}+XN_3\phi_{(3)}\phi_{(-3)}+ \nonumber \\ &&+M_7\phi_{(7)}\phi_{(-5)}+M_5\phi_{(5)}\phi_{(-3)}+M_3\phi_{(3)}\phi_{(-1)}\label{wnof}
\end{eqnarray}

Now we have to study R-symmetry breaking in this model. All parameters can be chosen real and positive. The condition $|M^{-2}fN|\ll 1 $ is generally sufficient to avoid tachyonic directions for small $X$, so we choose $f/M_5^2$ to be small. 

\begin{figure}[htb!]
\begin{center}
\includegraphics[width=12cm,height=10cm]{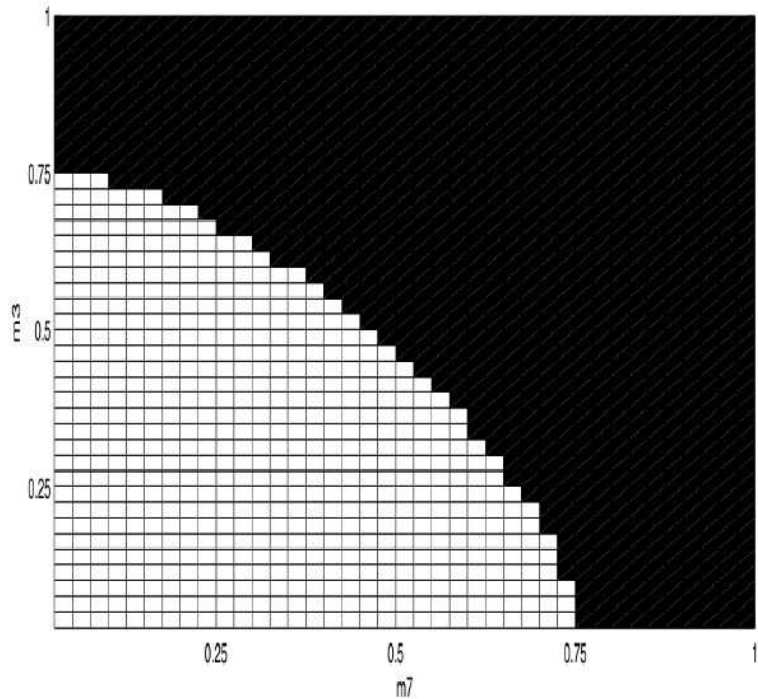}
\caption{The white area is the region of the plane $(m_7/m_5,m_3/m_5)$ where there is spontaneous R-symmetry breaking for $N_3=N_5=1$ and $f/M_5^2=0.001$.}
\label{figura}
\end{center}
\end{figure}

\begin{figure}[htb!]
\begin{center}
\includegraphics[width=12cm,height=10cm]{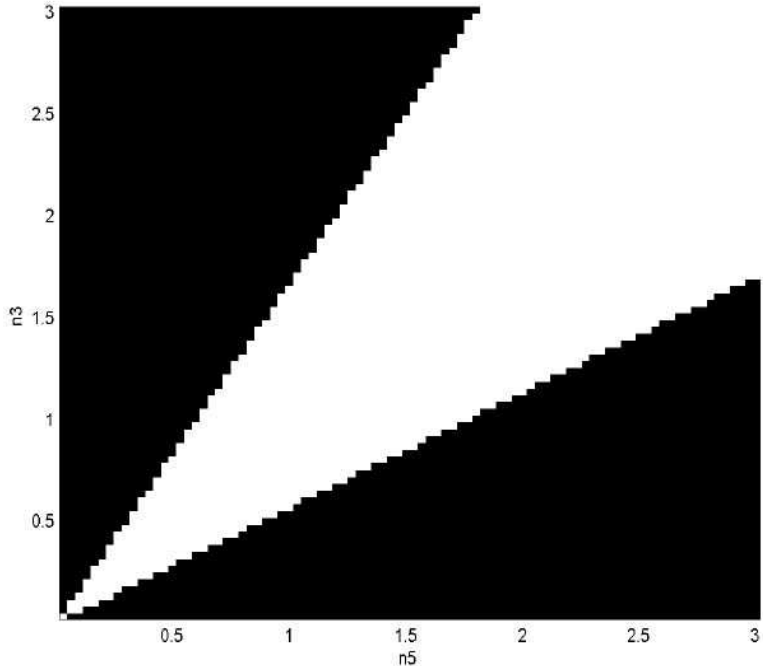}
\caption{The white area is the region of the plane $(n5,n3)$ where there is spontaneous R-symmetry breaking for $M_3/M_5=0.2,M_7/M_5=0.3$ and $f/M_5^2=0.001$. (We thank M. Cortelezzi for collaboration.)}
\label{figura2}
\end{center}
\end{figure}

 Numerical minimization of the Coleman-Weinberg potential for the model (\ref{wnof}) shows that there is spontaneous R-symmetry breaking in some 
 region of the parameter space, in particular for $N_3\sim N_5$ and $M_3,M_7<M_5$, as can be seen in figure \ref{figura},\ref{figura2}. 

\begin{figure}[htb!]
\begin{center}
\includegraphics{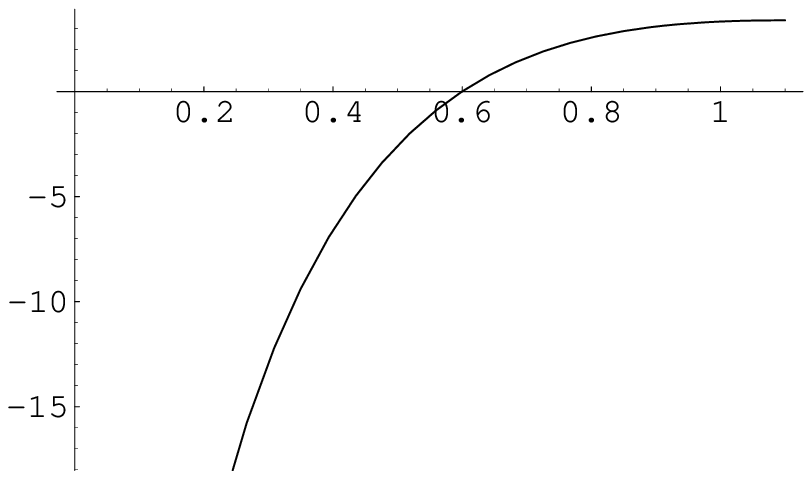}
\caption{Plot of $m_X^2M_5^2/f^2$ as a function of $M_3/M_5$.}
\label{pot2}
\end{center}
\end{figure}

\begin{figure}[htb!]
\begin{center}
\includegraphics{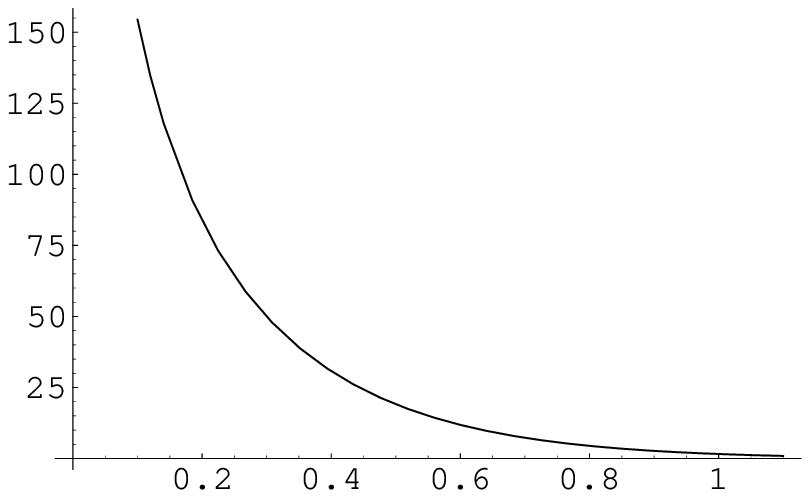}
\caption{Plot of $\lambda_XM_5^4/f^2$ as a function of $M_3/M_5$.}
\label{pot4}
\end{center}
\end{figure}

It is useful to show analytically that R-symmetry breaking happens in this region.  It is possible to expand  the Coleman-Weinberg potential at lowest order in $|\hat{M}^{-2}f\hat{N}|$ and $X$ and confirm the numerical results. The potential has the form $V(X)=V_0+m_X^2 |X|^2+\lambda_X|X|^4+\ldots$. In figure \ref{pot2} \ref{pot4} we plot the expressions found for $m_X^2M_5^2/f^2,\lambda_XM_5^4/f^2$ as functions of $M_3/M_5$ in the case $M_3=M_7,N_3=N_5=1$ and $f/M_5^2 \ll 1$.

We have studied the simplest model with complex representations, but we can also consider models with more fields. The results coming from numerical minimization are the same: these models have metastable quantum vacua which break R-symmetry for some range of parameters. 

In models with more pseudomoduli the range of parameters for spontaneous R-symmetry breaking becomes wider, because a linear combination of $X$ and $Y_a$ which acquires a negative $m^2$ is a sufficient condition for R-symmetry breaking. 
Numerical studies indicate that there are stable vacua which break R-symmetry in 
a large fraction of the parameter space for parameters $N_{ij},M_{ij}/M$ of order $O(1)$ and small $f/M$ \cite{mikk}. Non-hierarchical spontaneous R-symmetry breaking seems therefore a common feature of these models: this opens interesting possibilities for realistic model building. 

It would be interesting to explore the possibility of direct mediation of SUSY breaking using the model (\ref{wf}). This model could be made natural as in \cite{dinefs}, while the global symmetry could be gauged and then mediate SUSY breaking. Convincing models of direct mediation appeared for example in \cite{med1},\cite{med2} based on the ISS model \cite{iss1}. It could be possible to obtain similar results with some models of this section.

\subsection*{Acknowledgements:}
We would like to thank S. Cremonesi, G. Della Sala for useful discussions, F. Benini for bringing our attention on R-symmetries and  M. Bertolini, V. Fini, A. Romanino for comments on the draft.

\appendix

\section{Models with more pseudomoduli}\label{more2}

We generalize the analysis of \cite{shih} to include models with more pseudomoduli $Y_a$. The trick used in \cite{shih} is to rewrite the potential (\ref{cw}) as 
\be{V_{eff}^{(1-loop)}=-\frac{1}{32\pi^2}\int_0^\infty dv \ v^5 \left(\frac{1}{v^2+\mathcal{M}_B^2}-\frac{1}{v^2+\mathcal{M}_F^2}\right)}

The terms in the Coleman-Weinberg potential which are quadratic in $X,Y_a$  can be written as
\begin{eqnarray}V_{quad}=\frac{1}{16\pi^2}\mathrm{Tr}\int_0^\infty dv \ v^3 \left[ \frac{1}{v^2+\hat{M}^2+f\hat{N}} \left(\hat{Y}^2 -\frac{1}{2}\{ \hat{M},\hat{Y}\}\frac{1}{v^2+\hat{M}^2+f\hat{N}}\{ \hat{M},\hat{Y}\}\right)\right. + \nonumber \\
- \left. \frac{1}{v^2+\hat{M}^2} \left(\hat{Y}^2 -\frac{1}{2}\{ \hat{M},\hat{Y}\}\frac{1}{v^2+\hat{M}^2}\{ \hat{M},\hat{Y}\}\right)\right]
\end{eqnarray}
where \be{\hat{M}=\begin{pmatrix} 0 & M^\dagger \\ M & 0\end{pmatrix},\hat{N}=\begin{pmatrix} 0 & N^\dagger \\ N & 0\end{pmatrix},\hat{Y}=\begin{pmatrix} 0 & (NX+Q^aY_a)^\dagger \\ NX+Q^aY_a & 0\end{pmatrix}}
 We consider the case of small $f$, because in this limit we can neglect the possibility of tachyonic directions of $\phi$ fields 
 in a large range of values of $X,Y_a$ around the origin of the flat directions. Then at the lowest nonzero order in $|\hat
{M}^{-2}f\hat{N}|$ this expression reduces to
   \be{V_{quad}= \frac{f^2}{32\pi^2}\mathrm{Tr}\int_0^\infty dv \ v^3 \left[\mathcal{M}_1(v)\mathcal{M}_1^\dagger(v)-\mathcal{M}_2(v)\mathcal{M}_2^\dagger(v)\right]}
 with
\be{\mathcal{M}_1(v)=\frac{1}{\sqrt{v^2+\hat{M}^2}}\left(\hat{N}\frac{\sqrt{2}v}{v^2+\hat{M}^2}\hat{Y}\right)\frac{1}{\sqrt{v^2+\hat{M}^2}}}
\be{\mathcal{M}_2(v)=\frac{1}{\sqrt{v^2+\hat{M}^2}}\left(\hat{N}\frac{\hat{M}}{v^2+\hat{M}^2}\hat{Y}+\hat{Y} \frac{\hat{M}}{v^2+\hat{M}^2}\hat{N}\right)\frac{1}{\sqrt{v^2+\hat{M}^2}}}
 
 after eliminating some terms which do not contribute to the trace. The two terms are generally of the same order, but the contribution of the first term is always positive, while the second term always gives a negative contribution. Choosing $Y_a=0$, this expression is consistent with the corresponding formula in \cite{shih}. 
  
 If this expression is negative for some choice of $(X,Y_a)=(x,y_a)$ then the classical vacuum $X=0,Y_a=0$ is unstable because the linear combination $\bar{x}X+\bar{y}_aY_a$ of these fields has negative $m^2$. In this case there can be an R-symmetry breaking vacuum along one of these tachyonic directions.
 
It is clear that in these models the range of parameters for spontaneous R-symmetry breaking is much bigger than in models with a single pseudomodulus. In fact there are many directions in field space $X,Y_a$  which can be tachyonic, including the original one $X\neq 0, Y_a=0$.

\section{Solvability of $R\geq 0$ equations}\label{proof}
In this appendix we prove that it is always possible to solve the system of equations (\ref{ex}),(\ref{ea+}),(\ref{ea0}).

First of all, note that if there is a solution $\phi'_i,X'$ to (\ref{ea+}),(\ref{ea0}) which satisfies $N^{ij}\phi'_i\phi'_j\neq 0$, the equation (\ref{ex}) can be solved by rescaling all fields $\phi'_i\rightarrow\rho\phi'_i$ by a factor $\rho=(-f/N^{ij}\phi'_i\phi'_j)^{1/2}$. Therefore we only have to prove that (\ref{ea+}),(\ref{ea0}) can be solved with $N^{ij}\phi'_i\phi'_j\neq 0$.

The set of  fields $\phi_i$ of a given model (\ref{wshih}) can be decomposed into minimal subsets in such a way that two fields belonging to different subsets cannot appear in the same equation or in the same term of the superpotential\footnote{For example, fields with even and odd R-charge belong to different subsets.}. Each field $\phi_{(r)}$ interacts with X and with fields $\phi_{(2-r)j},\phi_{(-r)j}$ only and each equation has the form
\be{N_{(r,-r)}^{ij}X\phi_{(r)j}+M_{(2+r,-r)}^{ij}\phi_{(2+r)j}=0}
involving X and two fields  whose R-charges differ by 2. Different subsets give different systems of equations with no fields in common, so we will work with fields belonging to a minimal subset only, and we will neglect all the fields belonging to other subsets.

Let's prove the theorem for the case in which R-charges can be chosen in such a way that no field has $R=0$ or $R=1$. (We can always redefine R-charges by adding charges of $U(1)$ global symmetries.) First of all, note that it is always possible to choose an R-charge assignment so that all fields have integer R-charge. In fact if R-charges are not integer it is sufficient to consider the highest one $R_{max}$ and redefine them in the following way: $R(\varphi)\rightarrow \lceil R(\varphi)\rceil$ if $R(\varphi)-R_{max}$ is an even integer, $R(\varphi)\rightarrow \lfloor R(\varphi)\rfloor$ otherwise. A field with $R(\varphi)-R_{max}$ even is coupled only with fields with $R(\varphi)-R_{max}$ not even, therefore
 this defines a consistent R-charge assignment with only integer R-charges.
 
If there are no fields with $R=0$ or $R=1$,  we have a set of fields of $2m$ different R-charges
 $\phi_{(k)j}$, $\phi_{(2+k)j}$ \ldots $\phi_{(2m+k)j}$ and $\phi_{(2-k)j}$, $\phi_{(-k)j}$ \ldots $\phi_{(2-2m-k)j}$ with integers $k,m$ satisfying $k>2,m>1$.  
Every term in the superpotential couples fields with R-charges of opposite sign, therefore there is an accidental U(1) symmetry whose charge is $S(\phi_i)=\mathrm{sign}(R(\phi_i))$. Using this symmetry, we redefine the R-symmetry to obtain $\phi^+_{(-1)j}$, $\phi^+_{(1)j}$ \ldots $\phi^+_{(2m-1)j}$ and $\phi^-_{(3)j}$, $\phi^-_{(1)j}$ \ldots $\phi^-_{(-2m+3)j}$ and the equations (\ref{ea+}),(\ref{ea0}) become as follow:
\begin{eqnarray}
&&N_{(-2m+3,2m-3)}^{ij}X\phi^+_{(2m-3)j}+M_{(-2m+3,2m-1)}^{ij}\phi^+_{(2m-1)j}=0\nonumber\\
&&N_{(-2m+5,2m-5)}^{ij}X\phi^+_{(2m-5)j}+M_{(-2m+5,2m-3)}^{ij}\phi^+_{(2m-3)j}=0\nonumber\\
&&\ldots\nonumber\\
&&N_{(1,-1)}^{ij}X\phi^+_{(-1)j}+M_{(1,1)}^{ij}\phi^+_{(1)j}=0\nonumber\\
&&N_{(1,-1)}^{ji}X\phi^-_{(1)j}+M_{(3,-1)}^{ji}\phi^-_{(3)j}=0\nonumber\\
&&N_{(-1,1)}^{ji}X\phi^-_{(-1)j}+M_{(1,1)}^{ji}\phi^-_{(1)j}=0\label{e+-}
\end{eqnarray}
where $N_{k,k'}^{ij}$ couples $\phi^-_{(k)i}$ and $\phi^+_{(k')j}$ and the same happens for $M_{k,k'}^{ij}$.

We have two systems of equations containing  $\phi^+$ and $\phi^-$ fields respectively. For each fixed value of $X, \phi^-_{(-1)j},\phi^+_{(-1)j}$ we  have two linear systems of $n^+,n^-$ equations in $n^+,n^-$ variables, which can always be solved provided that the related linear operators have nonzero determinants. This condition is verified because these determinants are products of $\det(M_{(2-k,k)})$ and these cannot be zero because $\det (M)=\prod_k\det(M_{(2-k,k)})\neq 0$. 
If we choose $\phi^-_{(-1)j},\phi^+_{(-1)j}$ to be different from zero\footnote{The requirements here and in the other cases should be stated more precisely. For example, these fields have to be chosen such that they do not belong to the kernel of the matrices $N_{(-1,1)},N_{(1,-1)}$ respectively. However similar conditions are easily satisfied for generic nonzero fields.}
, then also $\phi^-_{(1)j},\phi^+_{(1)j}$ are nonzero and generically $N^{ij}\phi_i\phi_j\neq 0$. This completes the proof of this case.

Now we will prove the theorem for the case with $\phi_{(1)}$. The equations (\ref{ea+}),(\ref{ea0}) become:
\begin{eqnarray}
&&N_{(2m-3,-2m+3)}^{ij}X\phi_{(2m-3)j}+M_{(2m-1,-2m+3)}^{ij}\phi_{(2m-1)j}=0\nonumber\\
&&N_{(2m-5,-2m+5)}^{ij}X\phi_{(2m-5)j}+M_{(2m-3,-2m+5)}^{ij}\phi_{(2m-3)j}=0\nonumber\\
&&\ldots\nonumber\\
&&N_{(-1,1)}^{ij}X\phi_{(-1)j}+M_{(1,1)}^{ij}\phi_{(1)j}=0
\end{eqnarray}
and, applying the same argument we used above, choosing $\phi_{(-1)j}\neq 0$ is a sufficient condition.
The case with $\phi_{(0)}$ is very similar, with equations:
\begin{eqnarray}
&&N_{(2m-2,-2m+2)}^{ij}X\phi_{(2m-2)j}+M_{(2m,-2m+2)}^{ij}\phi_{(2m)j}=0\nonumber\\
&&N_{(2m-4,-2m+4)}^{ij}X\phi_{(2m-4)j}+M_{(2m-2,-2m+4)}^{ij}\phi_{(2m-2)j}=0\nonumber\\
&&\ldots\nonumber\\
&&N_{(0,0)}^{ij}X\phi_{(0)j}+M_{(2,0)}^{ij}\phi_{(2)j}=0\nonumber\\
&&N_{(0,-2)}^{ij}X\phi_{(-2)j}+M_{(2,0)}^{ij}\phi_{(0)j}=0
\end{eqnarray}
and choosing $\phi_{(-2)j}\neq 0$ is enough.

To complete the proof, we must discuss what happens when there are abelian or non-abelian symmetries that constrain the form of $M,N$. The only difference is that now the equations are classified not only by their R-charge, but also by other charges. However this has no effect on the above arguments, provided that we consider systems of equations of the same charge\footnote{From another point of view, 
two fields whose charges are not equal or complex conjugate belong to different minimal subsets.}. The proof is complete.

\section{Conditions for runaway}\label{cond}
We discuss some conditions for the existence of runaway directions. We consider only the case of minimal subsets. 

If it is possible to solve all the equations with $R>0$ for a generic choice of fields $\phi_i$, then it is always possible to minimize $V^+$ (or $V^-$). If the minimum is zero, there is a SUSY runaway direction, otherwise there is a non-SUSY runaway vacuum. 

In models with no fields with $R=0,1$, this is possible if $n_Y\geq \frac{n_\phi}{2}+n_{(1)}^- -n_{(2m-1)}^+-1$. We consider the R-charge choice of appendix \ref{proof}. We can see the equations with $R>0$ as $\frac{n_\phi}{2}+n_{(1)}^-$ generic linear equations in $n_Y+1+n_{(2m-1)}^+$ variables, that can be solved if the above condition is satisfied.

In models with a field with $R=1$ it is possible to repeat the above argument and obtain the condition $n_Y \geq \frac{n_\phi}{2}+\frac{n_{(1)}}{2} -n_{(2m-1)}-1$. 

In models with a field with $R=0$ the argument is different, because in this case we need to solve also equations with $R=0$ which contain $X,Y_a$. Considering also these equations, we obtain the condition $n_Y \geq \frac{n_\phi}{2}+n_{(0)} -n_{(2m)}-1$. 

The above conditions imply SUSY or (generally) non-SUSY runaway vacua. To obtain conditions which imply SUSY vacua, we need to solve all the equations with $R\geq 0$. Consider the case with no fields with $R=0,1$. Solving all the equations with $R>0$, we end with a set of $n_Y$ equations with $R=0$ of the form $\sum_{k>0}\phi^+_{(-1)}P^a_{(-k)}\phi^-_{(-k)}=0$ where $P^a_{(-k)}$ are generic matrices which depend on $X,Y_a$. These equation have a nonzero solution (choosing  a generic nonzero $\phi^+_{(-1)}$) if $\frac{n_\phi}{2}-n^-_{(3)}\geq n_Y+1$, so the condition is $ n_Y \leq \frac{n_\phi}{2}-n^-_{(3)}-1$. 

Similar conditions can be found for the other cases. If there are fields with $R=1$ the condition is $n_Y \leq \frac{n_\phi}{2}-\frac{n_{(1)}}{2} -n_{(-1)}-1$, while if there are fields with $R=0$ the condition is $n_Y \geq \frac{n_\phi}{2}-n_{(0)}-1$.


\section{SUSY vacua remnants}\label{vacua}
The existence of an R-symmetry is a sufficient condition for SUSY breaking in the models discussed in sections \ref{simple},\ref{more}. More generally, R-symmetry is a necessary condition for SUSY breaking under some condition of genericity of the superpotential. 

Consider a superpotential $W(\varphi_a)$ which has an R-symmetry and breaks SUSY spontaneously, and additional terms $W^r_{R \!\!\!\! \slash}(\varphi_a)$ which does not have R-charge 2. An immediate consequence of the statements above  is that the theory defined by
\be{W_\varepsilon=W+\varepsilon_r W_r^{R \!\!\!\! \slash}}
generally has supersymmetric vacua $\tilde{\varphi}_a(\varepsilon)$ which satisfy \be{\partial_b W_\varepsilon(\tilde{\varphi}_a(\varepsilon))=\partial_b W(\tilde{\varphi}_a(\varepsilon))+\varepsilon_r \partial_b W_r^{R \!\!\!\! \slash}(\tilde{\varphi}_a(\varepsilon))=0}
so the SUSY-breaking vacua which survive for $\varepsilon_r\ll 1$ are metastable. However, in the limit $\varepsilon_r\rightarrow 0$ the SUSY vacua are pushed to infinity.

The potential of the original theory along the direction of the SUSY vacua is \be{V(\tilde{\varphi}_a(\varepsilon))=\sum_b|\partial_b W(\tilde{\varphi}_a(\varepsilon))|^2=\sum_b|\varepsilon_r\partial_b W_r^{R \!\!\!\! \slash}(\tilde{\varphi}_a(\varepsilon))|^2}
Usually this potential doesn't vanish for $\varepsilon_r\rightarrow 0$ because  the contribution of $\partial_b W_r^{R \!\!\!\! \slash}(\tilde{\varphi}_a(\varepsilon))$ can grow as $1/\varepsilon_r$ or faster, so the  theory with $\varepsilon_r= 0$ has no memory of  SUSY vacua when they are pushed to infinity. 

However there is an interesting exception. If the condition
\be{\mathrm{sign}(R(\varepsilon_r))=\mathrm{sign}(R(\varepsilon_r'))=\mathrm{sign}(2-R(\varphi_b)) \quad \forall r,r' \ \mathrm{and}\ \forall \varphi_b \in W^{R \!\!\!\! \slash}}
 is satisfied, then the limit $\varepsilon_r\rightarrow 0$ can be interpreted as a rescaling with respect to the R-charges. In this case metastability of the R-symmetric superpotential can be easily explained, because the runaway vacuum is exactly the SUSY vacuum pushed to infinity as $\varepsilon_r\rightarrow 0$, and the runaway direction can be found following  the positions of SUSY vacua for $\varepsilon_r\neq 0$.

For the models with a single pseudomodulus there is a simple R-breaking perturbation which explains the metastability of vacua with $\phi=0$:
\be{W^{R \!\!\!\! \slash}=\sum_{R(\phi_j)> 2}\nu_{j} \phi_j}
This perturbation generates a SUSY vacuum with $|\phi|\sim 1/\nu$ that becomes a runaway vacuum when $\nu\rightarrow 0$.
A similar perturbation explains also the metastability of many vacua in models with more pseudomoduli.



\begin{thebibliography}{30} 

\bibitem{oraif}
  L.~O'Raifeartaigh,
  Nucl.\ Phys.\  B {\bf 96} (1975) 331.

\bibitem{iss1}
  K.~Intriligator, N.~Seiberg and D.~Shih,
  JHEP {\bf 0604} (2006) 021
  [arXiv:hep-th/0602239].

\bibitem{shih}
  D.~Shih,
  arXiv:hep-th/0703196.

\bibitem{med1}
  R.~Kitano, H.~Ooguri and Y.~Ookouchi,
  Phys.\ Rev.\  D {\bf 75} (2007) 045022
  [arXiv:hep-ph/0612139].


\bibitem{med2}
  C.~Csaki, Y.~Shirman and J.~Terning,
  arXiv:hep-ph/0612241.


\bibitem{ak}
  S.~A.~Abel and V.~V.~Khoze,
  arXiv:hep-ph/0701069.
\bibitem{mn1}
  H.~Murayama and Y.~Nomura,
  Phys.\ Rev.\ Lett.\  {\bf 98} (2007) 151803
  [arXiv:hep-ph/0612186].


\bibitem{as}
  O.~Aharony and N.~Seiberg,
  JHEP {\bf 0702} (2007) 054
  [arXiv:hep-ph/0612308].

\bibitem{mn2}
  H.~Murayama and Y.~Nomura,
  arXiv:hep-ph/0701231.



\bibitem{ns}
  A.~E.~Nelson and N.~Seiberg,
  Nucl.\ Phys.\  B {\bf 416} (1994) 46
  [arXiv:hep-ph/9309299].

\bibitem{is-lect}
  K.~Intriligator and N.~Seiberg,
  arXiv:hep-ph/0702069.


\bibitem{witten}
  E.~Witten,
  Phys.\ Lett.\  B {\bf 105} (1981) 267.


\bibitem{iss2}
  K.~Intriligator, N.~Seiberg and D.~Shih,
  arXiv:hep-th/0703281.


\bibitem{mikk}
M.~Cortelezzi, L.~Ferretti,
in preparation.



\bibitem{dinefs}
  M.~Dine, J.~L.~Feng and E.~Silverstein,
  Phys.\ Rev.\  D {\bf 74} (2006) 095012
  [arXiv:hep-th/0608159].

\end{thebibliography}

\end{document}